\documentclass[pra,aps,amsfonts,amsmath,superscriptaddress,twocolumn,showpacs,floatfix]{revtex4}
\usepackage{graphicx} 
\usepackage{epsfig} 

\begin{document}

\title{Effect of pairing correlations on incompressibility and 
symmetry energy in nuclear matter and finite nuclei}

\author{E. Khan}
\affiliation{Institut de Physique Nucl\'eaire, Universit\'e Paris-Sud, IN2P3-CNRS, F-91406 Orsay Cedex, France}
\author{J. Margueron}
\affiliation{Institut de Physique Nucl\'eaire, Universit\'e Paris-Sud, IN2P3-CNRS, F-91406 Orsay Cedex, France}
\author{G. Col\`o}
\affiliation{Dipartimento di Fisica, Universit\`a degli Studi and 
INFN Sez. di Milano, Via Celoria 16, 20133 Milano, Italy}
\author{K. Hagino}
\affiliation{Department of Physics, Tohoku University, Sendai, 980-8578, Japan}
\author{H. Sagawa}
\affiliation{Center for Mathematics and Physics, University of Aizu, Aizu-Wakamatsu, 965-8580 Fukushima, Japan}

\begin{abstract}
The role of superfluidity in the incompressibility and in the symmetry
energy is studied in nuclear matter and finite nuclei. Several pairing
interactions are used: surface, mixed and isovector dependent. Pairing has a
small effect on the nuclear matter incompressibility at saturation density,
but the effects are significant at lower densities. The pairing effect on
the centroid energy of the isoscalar Giant Monopole Resonance (GMR) is also
evaluated for Pb and Sn isotopes by using a microscopic constrained-HFB
approach, and found to change at most by 10 \% the nucleus incompressibility
$K_A$. It is shown by using the Local Density Approximation (LDA) that 
most of the pairing effect on the GMR centroid come from the low-density
nuclear surface.  
\end{abstract}

\pacs{21.10.Re, 21.65.-f, 21.60.Jz}

\date{\today}

\maketitle

\section{Introduction}

The nuclear incompressibility and the symmetry energy are closely related to
the isoscalar Giant Monopole Resonance (GMR)~\cite{blai80,colo04} and to the
isovector Giant Dipole Resonance (GDR)~\cite{trip08}, respectively.
The question of the effect of pairing correlations on the centroid
energy of the GMR has been first addressed in Ref. \cite{civ91}, and
has recently known a renewed interest ~\cite{jli08,khan09}. In both the Sn and
Pb isotopic chains, a specific increase of the GMR energy, associated with 
the corresponding finite nucleus incompressibility $K_A$, has been predicted
for the doubly magic $^{132}$Sn and $^{208}$Pb nuclei \cite{khan09,khan09b}.
A part of this apparent stiffness of doubly magic nuclei may be related to
pairing effects which in fact decrease the GMR energy in open-shell nuclei.
However, this study has been undertaken only with a pure surface pairing
interaction. It is therefore relevant to analyze more systematically this
effect using various pairing functionals. It should be noted that the
surface versus mixed nature of the pairing interaction is still under
discussion. For instance, a recent systematic study based on the odd-even
mass staggering seems to slightly favor a surface type of pairing
interaction \cite{ber09}.

The apparent decrease of incompressibility in superfluid nuclei raises the
question about a possible similar effect in infinite nuclear matter: until
now, the nuclear matter incompressibility is evaluated by neglecting the
pairing part of the functional. However, considering results for finite
nuclei, the equations of state used for neutron stars and supernovae
predictions should take into account pairing effects in the calculation of
the incompressibility modulus. Therefore the question of the behavior of
$K_\infty$ with respect to the pairing gap is raised since it seems clear
from nuclear data that the finite nucleus incompressibility $K_A$ decreases
with increasing pairing gap \cite{khan09}. A similar study for nuclear
matter, as well as a more systematic study in finite nuclei, should be
undertaken. This is the goal of the present work. It should be also noted
that we will not consider the neutron-proton $T$=0 pairing channel since the
nuclei considered are far from N=Z. 

The density dependence of the symmetry energy is one of the most debated
issues in nuclear physics at present. In fact, this has relevant
implications (i) for nuclear structure, since it has an important effect on
the size of the neutron root-mean-square (r.m.s.) radius in neutron-rich
nuclei, (ii) for nuclear reactions, e.g., in intermediate energy heavy ion
collisions where the isospin distribution of the reaction products is
dictated by the density dependence of the symmetry energy, and obviously
(iii) for the description of neutron stars. Review papers have been devoted
to this topic \cite{Steiner,Li}. Empirical information on the symmetry 
energy can be obtained from various sources, none of them being so far
conclusive by itself. No measurement of the neutron skin is available which
is accurate enough to constrain the symmetry energy. The properties of the
isovector GDR, of the low-lying electric dipole excitations, and of the
charge-exchange spin-dipole strength have been suggested as constraints (see
e.g. \cite{klim07}). In addition, different model analysis of heavy-ion
collisions have been proposed as a test of the main trend of the symmetry
energy at densities below saturation. However, in none of these studies, to
our knowledge, the problem of the pairing effects on the symmetry energy has
been addressed.

In this work, the effects of the pairing correlations on the
incompressibility and on the symmetry energy are studied consistently in
nuclear matter and in finite nuclei. The effects coming from the
correlation energy associated with the pairing force are included. These
pairing effects are studied in Section II. In Section III, a Local Density
Approximation (LDA) approach to the problem is employed in order to
understand the connection between the effects in infinite matter and finite
systems: the $^{120}$Sn nucleus is used as a benchmark. Finally, in Section
IV, a microscopic study on the role of superfluidity in the
incompressibility of finite nuclei is undertaken, employing several pairing
interactions: surface, mixed and isovector dependent.

\section{Nuclear matter}
\label{sec:nm}

In this Section, we study the effects of the pairing correlations on the
incompressibility and the symmetry energy in nuclear matter.

\subsection{Energy density}

The nuclear energy density $\epsilon$ ($\epsilon/\rho=E/V$)
is the sum of the Skyrme part, $\epsilon_\mathrm{Skyrme}$,
that includes the kinetic energy \cite{chab98}, plus the pairing energy 
density,  
\begin{equation}
\epsilon=\epsilon_\mathrm{Skyrme}+\epsilon_\mathrm{pair}.
\label{eq:edens}
\end{equation}
Here 
\begin{equation}
\epsilon_\mathrm{pair}=-\frac{1}{2}(N_{n}\Delta_n^2+N_{p}\Delta_p^2) \; .
\label{eq:e-pair}
\end{equation}
In Eq. (\ref{eq:e-pair}), $\Delta_{\tau}$ is the pairing gap and 
$N_{\tau}$ is the density of states, given by 
$N_{\tau}=m_\tau^* k_{F\tau}/\pi^2\hbar^2$, with $\tau=n,p$.
The energy density $\epsilon$ is a function of the total density $\rho$ and  
of the asymmetry parameter $\delta=(\rho_n-\rho_p)/\rho$.
In the $T=1$ channel, several pairing interactions are defined by 
\begin{equation}
v_\mathrm{pair}^\mathrm{IS}(\vec{r},\vec{r}^\prime)=
v_0\left(1-\eta \left(\frac{\rho}{\rho_0}\right)^\alpha\right)\,\delta(\vec{r}-\vec{r}'),
\label{eq:pair-surf}
\end{equation}
as a function of the value of $\eta$ that can range from 0 (volume-type 
pairing) to 1 (surface-type pairing). In Eq. (\ref{eq:pair-surf})  
the parameter $\alpha$ is set to 1 and $\rho_0$ is taken as the saturation
density of symmetric nuclear matter throughout all the study; 
moreover, we adopt the parameters $\eta$=0.35 and 0.65 for the volume-surface mixed-type 
pairing interactions, and $\eta$=1.0 for the 
surface-type interaction. The values of $v_0$
in all these cases are adjusted, for each $\eta$, in such a way to obtain equivalent 
results for the two neutron separation energy in the Sn
isotopes by  HFB calculations with  the SLy5 parameter set.  The 
pairing cutoff energy is set at 60 MeV \cite{khan09c}. These values
of $v_0$ are given in Table \ref{tab:ispair}. In the following, these pairing interactions 
will be denoted as IS, because they depend on the isoscalar density.

\begin{table}[t]
\setlength{\tabcolsep}{.2in}
\renewcommand{\arraystretch}{1.5}
\caption{Strength $v_0$ (in~MeV$\cdot$fm$^3$) of the pairing interactions obtained 
in the case of various Skyrme functionals. 
The values of $v_0$ are adjusted, for each $\eta$, to obtain equivalent 
results to those of Ref. \cite{khan09c} 
for the two neutron separation energy in the Sn
isotopes by the HFB calculations with the parameter set SLy5. The energy cutoff for 
the pairing window is taken to be 60 MeV.}
\label{tab:ispair}
\centering
\begin{tabular}{cccc}
\hline \hline
                    & $\eta$=0.35 & $\eta$=0.65 &$\eta$=1.00  \\
\hline
SLy5 \cite{chab98}  & -285        & -390        & -670        \\
Sk255 \cite{agra05} & -265        & -390        & -600        \\
Sk272 \cite{agra05} & -265        & -390        & -600        \\
LNS \cite{cao06}    & -250        & -390        & -670        \\
\hline \hline
\end{tabular}
\end{table}

We have also considered pairing interactions having the isovector
density dependence with $\delta=(\rho_n-\rho_p)/\rho$
in addition to the isoscalar density dependence. 
The MSH interaction is defined  as~\cite{mar07}
\begin{eqnarray}
v_\mathrm{pair}^\mathrm{MSH}(\vec{r},\vec{r}')&=&
v_0\left[1-(1-\delta)\eta_s \left(\frac{\rho}{\rho_0}\right)^{\alpha_s}
\right.\nonumber\\
&&\left.-\delta\eta_n \left(\frac{\rho}{\rho_0}\right)^{\alpha_n} \right]
\,\delta(\vec{r}-\vec{r}'),
\end{eqnarray}
with $v_0=-448$~MeV$\cdot$fm$^3$, $\eta_s=0.598$, $\alpha_s=0.551$, 
$\eta_n=0.947$, $\alpha_n=0.554$ (with a cutoff energy of 60~MeV).  
The YS interaction has also the isospin dependence 
as~\cite{yama08}
\begin{eqnarray}
v_\mathrm{pair}^\mathrm{YS}(\vec{r},\vec{r}')&=&v_0\left[1-\left(\eta_0 +\eta_1\tau_3\delta\right)
\frac{\rho}{\rho_0}
\right.\nonumber\\
&&\left.-\eta_2 \left(\delta\frac{\rho}{\rho_0}\right)^2
\right]\,\delta(\vec{r}-\vec{r}'),
\end{eqnarray}
with $v_0=-344$~MeV$\cdot$fm$^3$, $\eta_0=0.5$, $\eta_1=0.2$, $\eta_2=2.5$ (with a
cutoff energy of 50~MeV). The parameters mentioned have been used in 
connection with the SLy5 interaction. In the following these pairing interactions
will be denoted as IS+IV.

Effective interactions in the pairing channel are faced to double counting
problem \cite{coo59}. However using a zero range interaction constrained on
bare interactions allows to avoid this problem \cite{gar99}. This motivates
the use of a different interaction in the the pairing channel compared to
the particle-hole one. The pairing interactions used in this work are of two
types: either fitted on 
BCS gaps in symmetric and neutron matter calculated with bare interaction, 
 as in the MSH case, or designed to fit 
 observables in nuclei, as
in the IS and YS cases. These pairing interactions are featured with zero
range and with a cutoff, following the prescription of Ref. \cite{gar99}:
our aim is to provide reasonable pairing description in nuclei
 in order
to use the same interaction in nuclei and in nuclear matter. 
The MSH
interaction is considered as an extension of Ref. \cite{gar99} including
isospin dependence. It has been shown that it is possible to study the
surface properties of the pairing interaction using slabs of nuclear matter
\cite{bal03,pan06}. However, it should be mentioned that one usually deals 
with only the first order in the diagrammatic expansion of the many body
equations. Thus adjusting the pairing gap on the bare interaction is done as
this level and is therefore perfectible: there are screening effects which
are at next order for instance. Therefore adjusting the gaps on the bare
interaction is complementary to other ways such as 
constraining the pairing
interaction 
to fit the gaps in finite nuclei. We use both
approaches as explained above. 

In the present work, symmetric nuclear matter is studied as well as the
behavior of the symmetry energy. On this purpose, it should be
noted that the MSH interaction is obtained by constraining  
the neutron and proton gaps of the bare interaction in both symmetric and
pure neutron matter.  
The YS interaction is adjusted on several nuclei with different
isospin and there is an explicit isospin dependence. To perform a study in
asymmetric nuclear matter, the non linearity of the energy gap on the
isospin degree of freedom should be considered \cite{zha10} and may be
studied  in a forthcoming study.

The pairing gap in uniform matter is obtained from  the 
BCS gap equation~\cite{bcs57}
\begin{equation}
\Delta_k = \sum_{k'} -v_{kk'} \frac{\Delta_{k'}}{2E_{k'}}
\label{eq:gap}
\end{equation}
solved under the condition of the particle number conservation. 
In a given volume $V$ one assumes constant density  given by
\begin{equation}
\rho_\tau = \frac{2}{V} \sum_k \left( 1 - \frac{e_\tau(k)-\mu_\tau}{E_{k,\tau}} \right),
\end{equation}
where the quasiparticle energy is defined as $E_{k,\tau}=\sqrt{(e_\tau(k)-\mu_\tau)^2+
\Delta_{k,\tau}^2}$ , $e_\tau(k)$ is the single particle energy, 
and $\mu_\tau$ is the chemical potential. 
In Eq. \eqref{eq:gap}, $v_{kk'}$ is the pairing matrix element for  
the  plane waves, namely $\langle k\bar k \vert v \vert k'\bar k'\rangle$. 
Notice that in the case of the zero-range pairing interaction, the pairing gap 
$\Delta_k$ is independent of $k$. 

\begin{figure*}[htb]
\begin{center}
\includegraphics[width=0.80\linewidth]{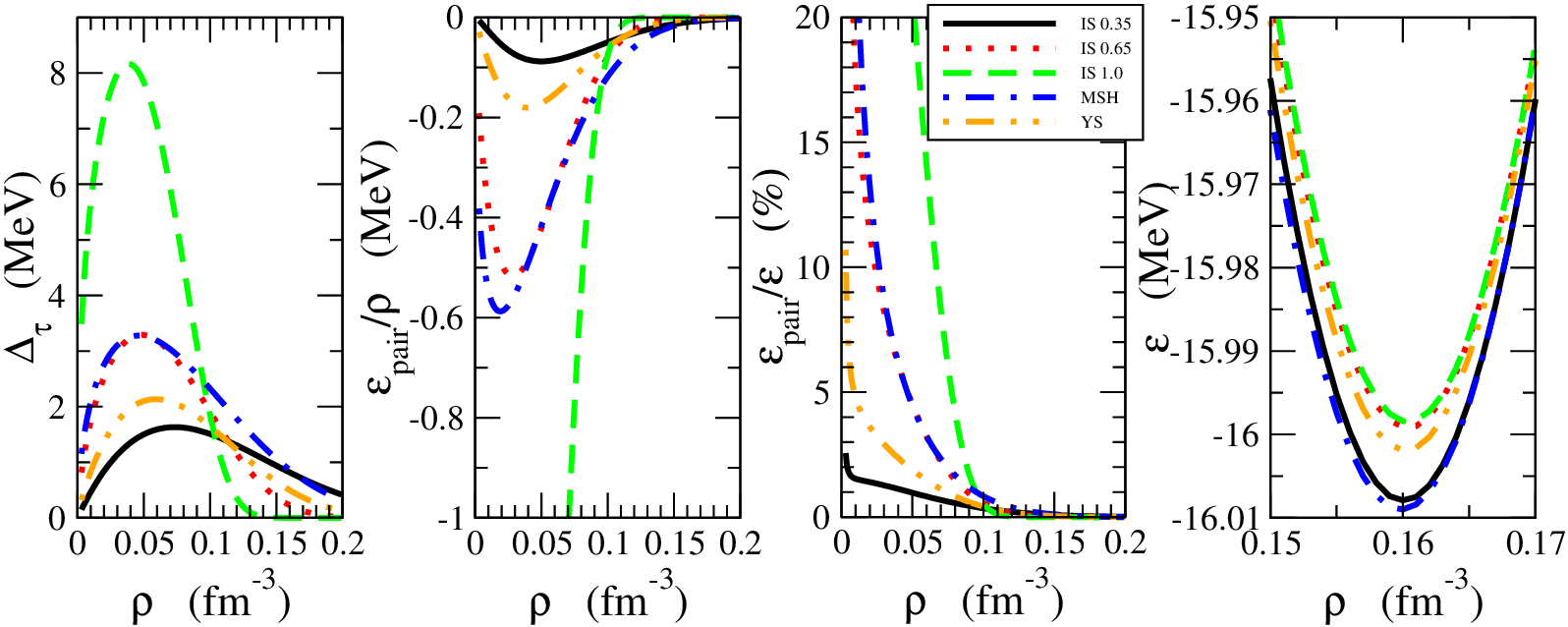}
\caption{(Color online) Pairing gap $\Delta_\tau$, pairing energy 
per particle $\epsilon_\mathrm{pair}/\rho$, percentage of pairing energy 
with respect to the total energy $\epsilon$ and equation of state around the
saturation point in symmetric matter obtained with various
pairing interactions employed in connection with the SLy5 Skyrme interaction.
 The solid, dotted and dashed lines correspond to the IS pairing interactions 
with $\eta=0.35, 0.65$ and 1.0, respectively, in Eq. (3). 
The dashed-dotted and dashed-dotted-dotted lines show the results of IS+IV 
interactions in Ref. [15] (MSH) and Ref. [16] (YS), respectively.
See the text for details.}
\label{fig:kinf1}
\end{center}
\end{figure*}

In Fig.~\ref{fig:kinf1} we display the pairing gap $\Delta_\tau$, the
pairing energy per particle $\epsilon_\mathrm{pair}/\rho$ and the percentage
of the pairing energy with respect to the total energy $\epsilon$ in
symmetric matter for the various pairing interactions together with the SLy5
Skyrme interaction in the mean field channel. There is a critical density
$\rho_c\approx$ 0.11~fm$^{-3}$ at which all the pairing interactions give 
almost the same result for the pairing gap. This has already been noticed in
Ref. \cite{khan09c} and may be related to the fact that in fitting the
two-neutron separation energy one is sensitive to the space region of the
nuclear surface, where the density is somewhat lower than the saturation
density: therefore the pairing gap is constrained rather at $\rho_c$ than at
$\rho_0$. Notice that the parameters of the MSH pairing interaction have been
determined without using constraints from finite systems, and $\Delta$ at
$\rho_0$ does not necessarily coincides with the one from the other
pairing interactions. Above $\rho_c$, the more surface-type the pairing interaction
(that is, the larger $\eta$ is taken), the smaller the pairing gap
$\Delta_\tau$. Below the critical density, the trend is reversed: the more
surface-type the pairing interaction is, the larger the pairing gap. The
contribution of the pairing energy is increased at low densities. Around the
saturation density, the pairing energy per particle is much smaller than the
binding energy ($-16$~MeV).

\subsection{Incompressibility and symmetry energy}

The compressibility $\chi$ is usually defined by 
\begin{equation}
\chi=-\frac{1}{V}\frac{\partial V}{\partial P}=
\frac{1}{\rho}\left(\frac{\partial P}{\partial \rho}\right)^{-1},
\label{eq:chi}
\end{equation}
and the pressure is related to the energy density $\epsilon$ by
\begin{equation}
P=-\frac{\partial E}{\partial V}=\rho^2\frac{\partial E/A}{\partial \rho}
=\rho\frac{\partial \epsilon}{\partial \rho}-\epsilon=\rho\mu-\epsilon.
\label{eq:pressure}
\end{equation}
The chemical potential is defined by
\begin{equation}
\mu=\frac{\partial E}{\partial A}=\frac{\partial \epsilon}{\partial \rho}.
\end{equation}
>From Eqs.~(\ref{eq:chi}) and (\ref{eq:pressure}), we obtain
\begin{equation}
\frac{1}{\chi}=\rho^2\frac{\partial^2 \epsilon}{\partial \rho^2}.
\end{equation}
The compressibility $\chi(\rho)$ is a function of the density $\rho$ and the asymmetry
parameter $\delta$. Furthermore one defines the incompressibility at the 
saturation density in symmetric nuclear matter by
\begin{equation}
K_\infty = \left.k_F^2 \frac{\partial^2 E/A}{\partial k_F^2}\right|_{k_{F\infty}}
=\left.9 \rho_0^2 \frac{\partial^2 E/A}{\partial \rho^2}\right|_{\rho_0}.
\label{eq:bulkmodulus}
\end{equation}
The relation between $K_\infty$ and $\chi$ is given by 
\begin{equation}
K_\infty = \frac{9}{\rho_0 \chi(\rho_0)}.
\label{eq:kinfchi}
\end{equation}
It is worth keeping in mind that the incompressibility $K_\infty$ is defined only at 
the saturation density $\rho_0$, and in particular the relation (\ref{eq:kinfchi}) 
is valid only for $\rho=\rho_0$. Indeed, more generally, we can define the 
density-dependent incompressibility (or bulk modulus) as \cite{fw71}
\begin{equation}
K(\rho)=\frac{9}{\rho \chi}=9\rho^2\frac{\partial^2 E/A}{\partial
\rho^2} + \frac{18}{\rho} P
\label{eq:krho}
\end{equation}
which coincides with the incompressibility $K_\infty$ at the saturation density. 

The symmetry energy $S(\rho$) is defined by
\begin{equation}
S(\rho) = \left.\frac{1}{2}\frac{\partial^2\epsilon/\rho}{\partial \delta^2}\right|_{\delta=0}.
\label{eq:symm}
\end{equation}
The symmetry energy can be expanded, around the saturation density, as 
\begin{equation}
S(\rho) = J+L\left(\frac{\rho-\rho_0}{3\rho_0}\right)+\frac{1}{2}K_\mathrm{sym} 
\left(\frac{\rho-\rho_0}{3\rho_0}\right)^2,
\label{eq:srho}
\end{equation}
where $J$ is defined by $J=S(\rho_0)$, $L=3\rho_0\frac{\partial S}
{\partial \rho}\vert_{\rho_0}$,
and $K_\mathrm{sym}=9\rho_0^2\frac{\partial^2 S}{\partial \rho^2}
\vert_{\rho_0}$.

\subsection{Results}

\begin{figure*}[htb]
\begin{center}
\includegraphics[width=0.80\linewidth]{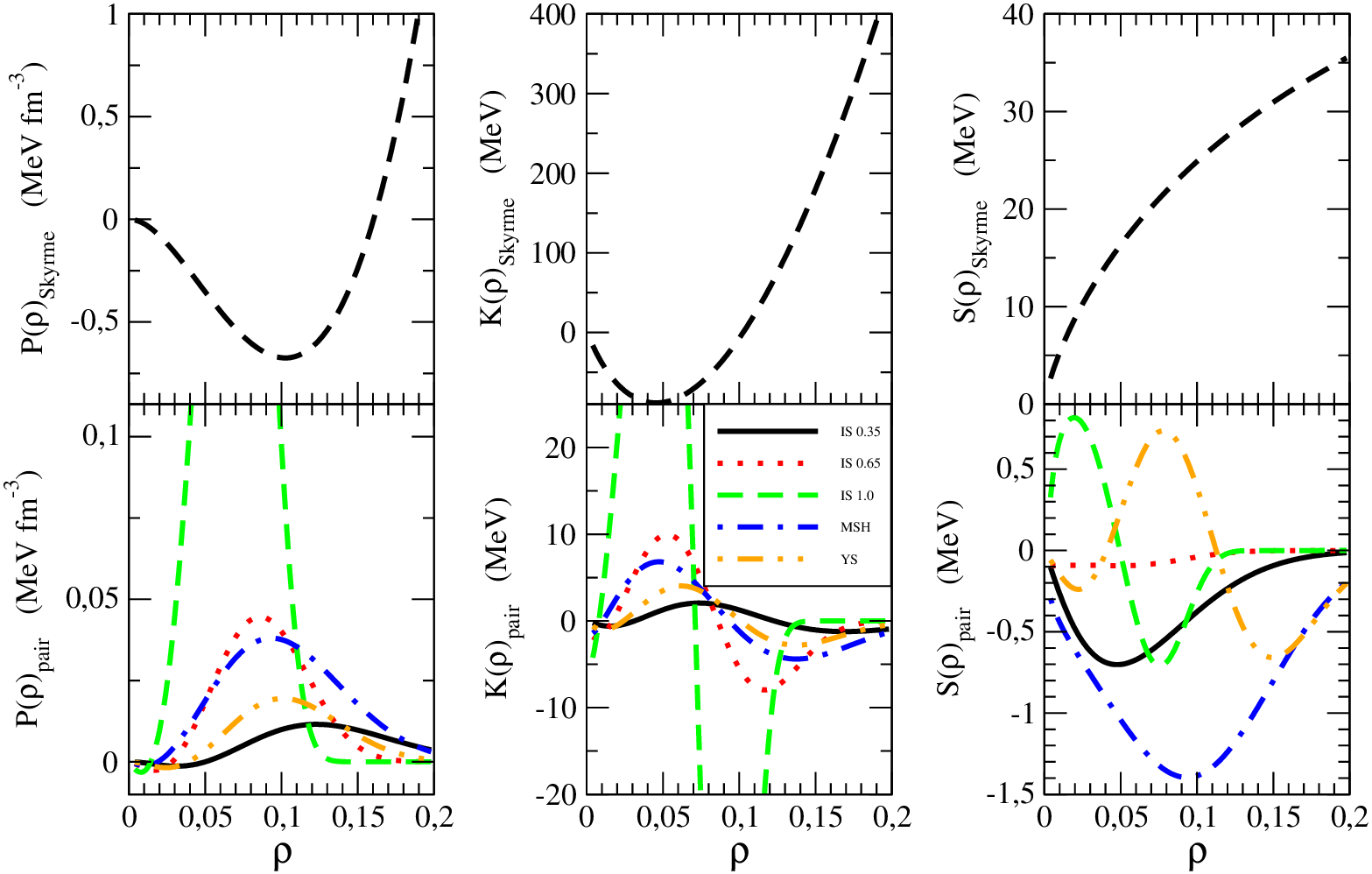}
\caption{(Color online) Pressure, incompressibility, and symmetry energy without pairing (top panels), and the
contributions of various pairing interactions 
 to these quantities (bottom panels).
The HF energy $\epsilon_{Skyrme}$ is calculated using the SLy5 interaction.
For details, see the caption to Fig. 1 and the text.}
\label{fig:kinf2}
\end{center}
\end{figure*}

Fig.~\ref{fig:kinf2} displays the pressure $P$ in Eq.~(\ref{eq:pressure}),
the incompressibility $K(\rho)$ in Eq.~(\ref{eq:krho}), and the symmetry
energy $S(\rho)$ in Eq.~(\ref{eq:srho}) without pairing (top panels), and
the contribution of pairing to these quantities (bottom panels), using the
SLy5 interaction. This contribution is calculated with the same equations,
but considering only the pairing term of the energy density in
Fig.~\ref{eq:edens}. The same pairing interactions have been considered
here as in Fig.~\ref{fig:kinf1}. Close to the saturation density, the
contribution from pairing is very small. This is also illustrated in
Table~\ref{tab:sat}: the pairing interaction has small effects at the
saturation density. In the case of the incompressibility $K_\infty$,
pairing can still produce a few \% effect (for instance $K_\infty$ is
changed from 230.2 MeV to 223.9 MeV in the case of the MSH pairing
interaction). The MSH, YS, IS 0.35 pairing interactions modify the
incompressibility by 3 to 6~MeV, that is, by about 2\%. It should be noted
that at the saturation density, the contribution to the slope parameters of
the symmetry energy, $L$, and $K_\mathrm{sym}$, of the interactions MSH and
YS is larger than that of the other IS forces. The effects on $L$ can be
about 15\% while $K_\mathrm{sym}$ can be modified in an important way. This
is related to the dependence of these pairing interaction on the isovector
density.

\begin{table}[t]
\setlength{\tabcolsep}{.05in}
\renewcommand{\arraystretch}{1.5}
\caption{Properties of nuclear matter for various pairing interactions.
The SLy5 Skyrme force is used for the mean field. 
}
\label{tab:sat}
\centering
\begin{tabular}{ccccccc}
\hline \hline
Pairing & $\rho_0$ & E/A($\rho_0$) & $K_\infty$ & $J$ & $L$ & $K_\mathrm{sym}$ \\
& [fm$^{-3}$] & [MeV] & [MeV] & [MeV] & [MeV] & [MeV] \\
\hline \hline
no pairing & 0.1604 & -15.999  & 230.2 & 32.03 & 48.25  & -112.3 \\\hline
IS  $\eta$=0.35  & 0.1601 & -15.998 & 227.3 & 31.93 & 48.49 & -129.7 \\
IS  $\eta$=0.65  & 0.1603 & -15.998 & 228.1 & 32.02 & 48.30 & -113.7 \\
IS  $\eta$=1.00  & 0.1604 & -15.999 & 230.1 & 32.03 & 48.25 & -112.3 \\
MSH      & 0.1599 & -15.998 & 223.9 & 31.33 & 55.77 & -139.7 \\
YS       & 0.1602 & -15.998 & 227.0 & 31.39 & 52.04 & 13.2 \\
\hline \hline
\end{tabular}
\end{table}

However at lower densities, the pairing effects become appreciably larger as
seen in Fig.~\ref{fig:kinf2}. In the case of the pure surface pairing, there
are important contributions to the pressure, incompressibility and symmetry
energy: these quantities can be strongly affected by pairing, which can lead
to variations up to about a factor 2. Other pairing interactions also
provide significant corrections to the pressure and the incompressibility,
typically, around 10\%. In the case of the symmetry energy, below
$\rho\approx 0.1$~fm$^{-3}$, the IS+IV pairing interactions (MSH and YS) 
predict an opposite and positive contribution compared to the negative
contributions of IS pairing interactions. It should be noted that the
pairing contribution to these quantities is generally larger at densities
below saturation.

To obtain a more general view of the pairing effect on the
incompressibility, Table \ref{tab:kinf} displays the $K_\infty$ values
obtained for SLy5, LNS, Sk255 and Sk272 Skyrme functionals, with various
pairing interactions. The correction induced by the pairing interaction IS
0.35 is largest one among the IS interactions and reduces the
incompressibility $K_\infty$ by about 3 MeV. The MSH interaction induces a
correction of 6.3 MeV on the incompressibility (Table \ref{tab:sat}). It
should be noted that the pure surface pairing interaction provides no
modification of $K_\infty$. Depending on the Skyrme models, there shall also
be an effect due to the different effective masses $m^*/m$, but they are
incorporated in the renormalization of the pairing interaction parameter
$v_0$ shown in Table~\ref{tab:ispair}.

\begin{table}[t]
\setlength{\tabcolsep}{.05in}
\renewcommand{\arraystretch}{1.5}
\caption{Nuclear matter incompressibility $K_\infty$ (MeV) for SLy5
\cite{chab98}, LNS \cite{cao06}, Sk255 \cite{agra05} and Sk272 \cite{agra05} Skyrme
functionals. The dependence of $K_\infty$ on the pairing interaction is displayed: 
mixed (IS  $\eta$=0.35), surface (IS  $\eta$=1.00).}
\label{tab:kinf}
\centering
\begin{tabular}{ccccc}
\hline \hline
Pairing & SLy5 & LNS & Sk255 & Sk272 \\
\hline\hline
no pairing & 230.2 & 211.0 & 255.2 & 271.8 \\\hline
IS  $\eta$=0.35 & 227.3 & 208.4 & 251.3 & 268.3 \\
IS  $\eta$=1.00 & 230.1 & 211.0 & 255.2 & 271.8 \\
\hline \hline
\end{tabular}
\end{table}

It is expected that the above pairing effects at low densities may also 
affect finite nuclei. In the case of incompressibility, we can define a
finite nucleus value $K_A$ and expect that this value is affected by the 
pairing more than $K_\infty$, due to the presence of a lower density region,
i.e. the nuclear surface. We analyze this point in the next Section, and
we argue that a similar reasoning holds for the symmetry energy.

\section{Local Density Approximation}

This section relates the general expressions in uniform matter obtained in 
Sec. \ref{sec:nm} with the observables in finite nuclei in the local density
approximation. The aim is to estimate the role of pairing in the
incompressibility and symmetry energy of finite nuclei in a simple and
transparent way. The validity of the LDA will be estimated by comparing the
predicted nuclei incompressibility with the one obtained by a microscopic
approach.

\begin{figure*}[htb]
\begin{center}
\includegraphics[width=0.80\linewidth]{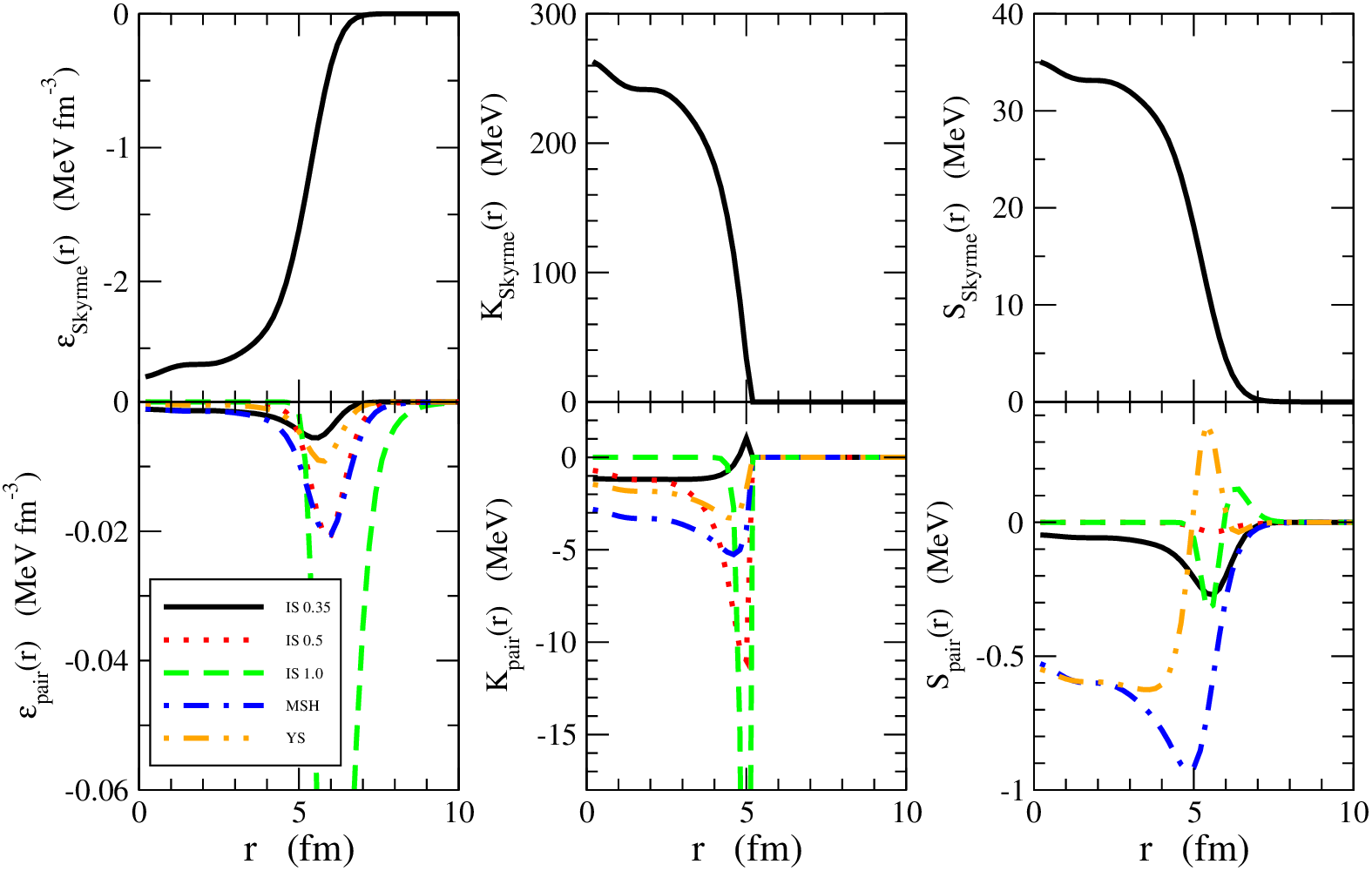}
\caption{(Color online) Radial dependence of $\epsilon(r)$, 
$K_{Nucl.}(r)$ and $S_{A}(r)$ for the various pairing interactions
considered (bottom panel) using the SLy5 force in $^{120}$Sn within the LDA.
The contribution of the Skyrme term
plus the kinetic term is displayed in the top panel. 
See the captions to Fig. 1 and the text for details.}
\label{fig:kinf3}
\end{center}
\end{figure*}

The binding energy per nucleon in the LDA reads
\begin{equation}
B_A(N,Z) = B_\mathrm{Nucl}(N,Z)+B_\mathrm{Coul}\frac{Z^2}{A^{4/3}}\;,
\end{equation}
where $B_\mathrm{Nucl}(N,Z)$ includes the bulk, surface and pairing contributions. 
It is defined by
\begin{equation}
B_\mathrm{Nucl}(N,Z) = \frac{1}{A} \int d^3r\ \epsilon(r),
\label{eq:mass1}
\end{equation}
where 
$\epsilon(r)=\epsilon\left(\rho_n(r),\rho_p(r)\right)=\epsilon_{Skyrme}(r)
+\epsilon_{pair}(r)$ as was defined in Eq. (1).  
The neutron and proton densities
$\rho_n(r),\rho_p(r)$ can be obtained, in the present context, by means of a spherical HF calculation.
The pairing contribution to the binding energy is defined by
\begin{equation}
B_\mathrm{pair}(N,Z) = \frac{1}{A} \int d^3r\ \epsilon_\mathrm{pair}(r).
\end{equation}

$B_\mathrm{Nucl}$ can be expanded around the saturation density,
\begin{equation}
B_\mathrm{Nucl}(N,Z)\approx B_\infty
+\frac{1}{2}K_A\left(\frac{\rho-\rho_0}{3\rho_0}\right)^2+S_{A}\delta^2 \; ,
\end{equation}
where the incompressibility in nuclei, $K_A$, is defined by 
$K_A=K_\mathrm{Nucl}+K_\mathrm{Coul}\cdot Z^2 A^{-4/3}$, and 
\begin{equation}
K_\mathrm{Nucl} = 9 \rho_0^2 \frac{\partial^2 B_\mathrm{Nucl}(N,Z)}{\partial \rho^2},
\label{eq:kalda}
\end{equation}
while the pairing contribution to the incompressibility is defined by
\begin{equation}
K_\mathrm{pair} = 9 \rho_0^2 \frac{\partial^2 B_\mathrm{pair}(N,Z)}{\partial \rho^2}.
\label{eq:kapair}
\end{equation}
The Coulomb contribution, $K_\mathrm{Coul}$, can be evaluated using for instance 
the Thomas-Fermi approximation [cf. Eq. (A1) in Ref.~\cite{sag07}]. It will not be 
included in the present work but the value obtained in Ref.~\cite{sag07} is $-$8 MeV 
$<$ $K_\mathrm{Coul}$ $<$ $-$4 MeV, depending on the interaction which is used.

The symmetry energy in finite nuclei, $S_{A}$, is defined by
\begin{equation}
S_{A} = \left.\frac{1}{2}\frac{\partial^2 B_\mathrm{Nucl}(N,Z)}{\partial \delta^2}\right|_{\delta=0},
\label{eq:as}
\end{equation}
and the contribution of the pairing correlations to $S_A$ is defined by
\begin{equation}
S_\mathrm{pair} = \left.\frac{1}{2}\frac{\partial^2 B_\mathrm{pair}(N,Z)}
{\partial \delta^2}\right|_{\delta=0}. 
\label{eq:as1}
\end{equation}

Introducing the mass formula~(\ref{eq:mass1}) into Eq.~(\ref{eq:kalda}), one obtains
\begin{equation}
K_\mathrm{Nucl} = \frac{\rho_0}{A} \int d^3r \; K_\mathrm{Nucl}(r)
\label{eq:ka2}
\end{equation}
with
\begin{equation}
K_\mathrm{Nucl}(r) = \frac{\rho_0}{\rho} \; K(\rho(r)).
\label{eq:knuclk}
\end{equation}

For small values of the density ($\rho \lesssim 0.6 \rho_0$, that is r
$\gtrsim$ 5 fm in $^{120}$Sn), the incompressibility is found to be 
negative: this is due to the spinodal instability in nuclear matter 
which is not present in finite systems \cite{mar03}. For this reason, the
integral (\ref{eq:ka2}) is limited to the region where $K_\mathrm{Nucl}(r)$ 
is positive. In this way, the spurious component due to the spinodal instability is
removed.

Introducing the quantity
\begin{equation}
S_{A}(r) = \left.\frac{1}{2 \rho_0}\frac{\partial^2 \epsilon}{\partial \delta^2}\right|_{\delta=0}
=\frac{\rho}{\rho_0} S(\rho(r)),
\label{eq:sar}
\end{equation}
the symmetry energy in nuclei~(\ref{eq:as}) reads
\begin{equation}
S_{A} = \frac{\rho_0}{A} \int d^3r \; S_{A}(r).
\label{eq:asa}
\end{equation}

We first perform a self-consistent HF calculation which provides the neutron
and proton densities in $^{120}$Sn. From these densities we deduce the
radial distributions of mean field part and pairing part of $\epsilon(r)$
given in Eq. (1) , $K_\mathrm{Nucl}(r)$ given in Eq. (\ref{eq:knuclk}), and
$S_{A}(r)$ given in Eq. (\ref{eq:sar}): these radial functions are shown in
Fig. \ref{fig:kinf3}. As expected from the results discussed in the previous
section, the pairing effects on $\epsilon(r)$, $K_\mathrm{Nucl}(r)$ and
$S_{A}(r)$ come from the low density surface region.

From Eqs.~(\ref{eq:mass1}), (\ref{eq:ka2}) and (\ref{eq:asa}), we obtain, in
the SLy5 case, 
$B_A=-$13.5~MeV, $K_\mathrm{Nucl}$=119.8~MeV, and $S_{A}$=25.7~MeV 
 without the contribution due to the pairing correlations. The
Coulomb contribution has not been included. The value for
$K_\mathrm{Nucl}$ should be compared with that of 141 MeV obtained by the
constrained HFB (CHFB)
calculations presented in the next section (cf. Fig. \ref{fig:kinfkasn2}). 
It should be noted that the CHFB
calculations take into account the contribution coming from the Coulomb
interaction. This contribution is estimated to be about 20 MeV in $^{120}$Sn,
using the values of $K_{Coul}$ from Ref. \cite{sag07}. The good agreement
between the LDA and the CHFB results ensures that LDA provides a sound framework 
to relate the nuclear matter incompressibility and the finite nucleus one.

The contributions of pairing correlations to the binding energy, the bulk
modulus and the symmetry energy are shown in Tab.~\ref{tab:120Sn} for the
various pairing interactions considered. The contribution of the
surface-type pairing (IS $\eta$=1.0) reduces $K_{A}$ by about 5\%, whereas,
for the IS mixed-type ($\eta$=0.35 or 0.65) and the IS+IV (MSH and YS)
pairing interactions, the effect on $K_A$ is predicted to be smaller. In
Table \ref{tab:120Sn}, it is also observed that pairing effects affect the
binding energy by few percents, up to 5\% for the surface-type pairing
interaction. For the symmetry energy, pairing effects are negligible, being
below 1\% except the IS+IV pairing (MSH). 

\begin{table}[t]
\setlength{\tabcolsep}{.2in}
\renewcommand{\arraystretch}{1.5}
\caption{Contributions of pairing correlations to the binding energy, the incompressibility
and the symmetry energy in $^{120}$Sn. The mean field is calculated 
by  using SLy5 interaction}
\label{tab:120Sn}
\centering
\begin{tabular}{cccc}
\hline \hline
Pairing  & $B_\mathrm{pair}$ & $K_\mathrm{pair}$ & $S_\mathrm{pair}$ \\
         & [MeV]             & [MeV]             & [MeV]             \\
\hline
IS  $\eta$=0.35  & $-$0.03           & $-$0.5            &  $-$0.25 \\
IS  $\eta$=0.65  & $-$0.11           & $-$3.9            &  $-$0.03 \\
IS  $\eta$=1.00  & $-$0.64           & $-$6.0            &  $-$0.03 \\
MSH      & $-$0.13           & $-$3.2            & $-$0.93  \\
YS       & $-$0.05           & $-$1.9            & $-$0.24  \\
\hline \hline
\end{tabular}
\end{table}

\section{Finite Nuclei}
\label{sec:CHFB}

In the previous section, the LDA has shown that the pairing effect on the
symmetry energy is negligible, whereas appreciable 
effects are observed in the case of the incompressibility. 
In this section, the role of pairing effects on the finite nucleus incompressibility $K_A$
is investigated using a microscopic approach.

We use the sum rule approach in order to calculate the centroid energy
of the isoscalar GMR. It is known that the finite nucleus incompressibility 
$K_A$ is related to that centroid energy by means of the relation
\begin{equation}
E_{\rm ISGMR}=\sqrt{\frac{\hbar^2K_A}{m \langle r^2 \rangle}},
\end{equation} 
where $m$ is the nucleon mass and $\langle r^2 \rangle$ denotes the
ground-state expectation value of the square radius. In a microscopic approach, for
the so-called scaling $K_A$, we calculate the energy as 
\begin{equation}
E_{\rm ISGMR}=\sqrt{\frac{m_1}{m_{-1}}}.
\end{equation} 
where the $k-th$ energy weighted sum rule is  defined as 
\begin{equation}
m_k=\sum_iE^k_i|\langle i|\hat{Q}|0\rangle |^2,
\end{equation}
with the RPA excitation energy  $E_i$  and the isoscalar monopole transition 
operator,
\begin{equation}
\hat{Q}=\sum_{i=1}^A r_i^2.
\end{equation}
The $m_1$ moment is evaluated by the  double commutator using the 
Thouless theorem \cite{tho61}:
\begin{equation}
m_1=\frac{2\hbar^2}{A} \langle r^2 \rangle. 
\end{equation}

In the present HFB calculations, the energy cutoff is 60 MeV, and the
$j_{\rm max}$ value is 15/2 in the case of IS pairing and extended to 25/2
for the IS+IV MSH pairing, which are the cutoffs used in the design of these
pairing interactions \cite{mar07,khan09c}, in order to ensure convergence of
the results. It should be noted that the strength $v_0$ of the IS pairing
interactions is adjusted in nuclei for its corresponding $j_{\rm max}$ and
energy cutoff. Therefore the different values of $j_{\rm max}$ between IS
and IS+IV pairing has little influence on the calculations.

Concerning the evaluation of the $m_{-1}$ moment, the CHFB approach 
is used. It should be noted that the extension of the constrained HF 
method \cite{boh79,colo04} to the CHFB case has been recently demonstrated in 
Ref. \cite{cap09} and employed also in \cite{khan09}. The CHFB Hamiltonian 
is built by adding the constraint associated with the IS monopole operator, namely 
\begin{equation}
\hat{H}_{constr.}=\hat{H}+\lambda\hat{Q}, 
\end{equation}
and the $m_{-1}$ moment is obtained from the derivative of the expectation value of 
the monopole operator on the CHFB solution $\vert\lambda\rangle$,
\begin{equation}
m_{-1}=-\frac{1}{2}\left[\frac{\partial}{\partial\lambda}\langle\lambda|
\hat{Q}|\lambda\rangle\right]_{\lambda=0}.
\end{equation}

\begin{figure}[tb]
\begin{center}
\scalebox{0.35}{\includegraphics{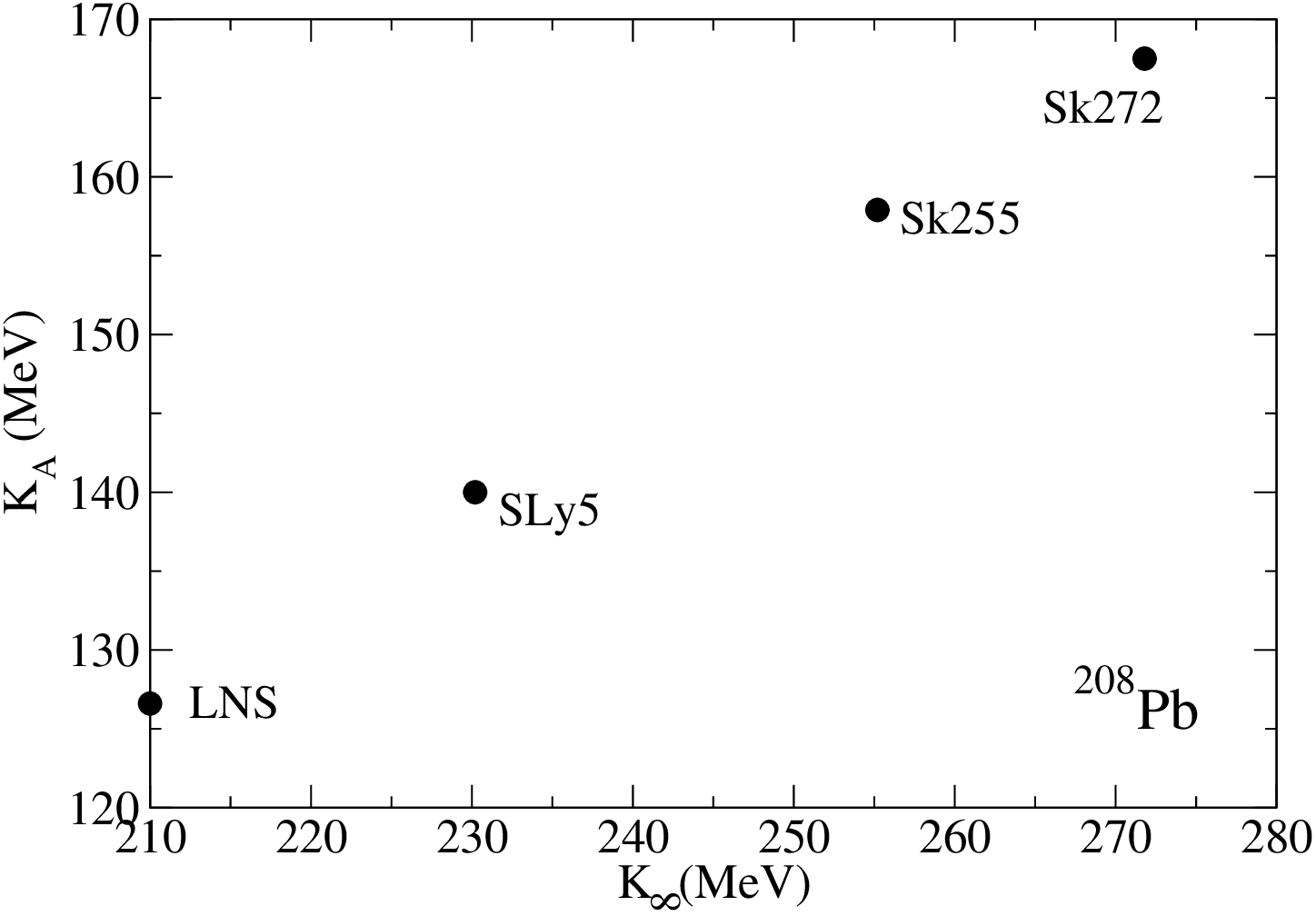}}
\caption{$K_\infty$ versus $K_A$ 
for $^{208}$Pb obtained by  the CHF method for several Skyrme interactions.}
\label{fig:kinfkapb}
\end{center}
\end{figure}

We first investigate the correlations between nuclear matter
incompressibility $K_\infty$ and finite nucleus, $K_A$, in the absence of
pairing correlations. This correlation has been found in all the previous
papers on the subject, but is reported here as a benchmark for further
considerations concerning the effect of pairing. For this purpose we choose
the doubly magic $^{208}$Pb nucleus. Figure \ref{fig:kinfkapb} displays
$K_A$ (obtained with the CHF method) versus $K_\infty$ for the four Skyrme
interactions LNS, SLy5, Sk255 and Sk272. These four interactions span a
large range of incompressibilities and have been fitted by using different
physics inputs: the neutron matter EOS from realistic forces in the case of
SLy5, Br\"uckner-HF calculations in nuclear matter in the case of LNS, and
the empirical properties of symmetric uniform matter plus a few binding
energies and charge radii of selected nuclei (the same which had been used
to fit some relativistic functionals like NL3) in the case of Sk255 and 
Sk272. In this sense, these interactions provide representative samples of
the Skyrme functionals.

\begin{figure}[tb]
\begin{center}
\scalebox{0.35}{\includegraphics{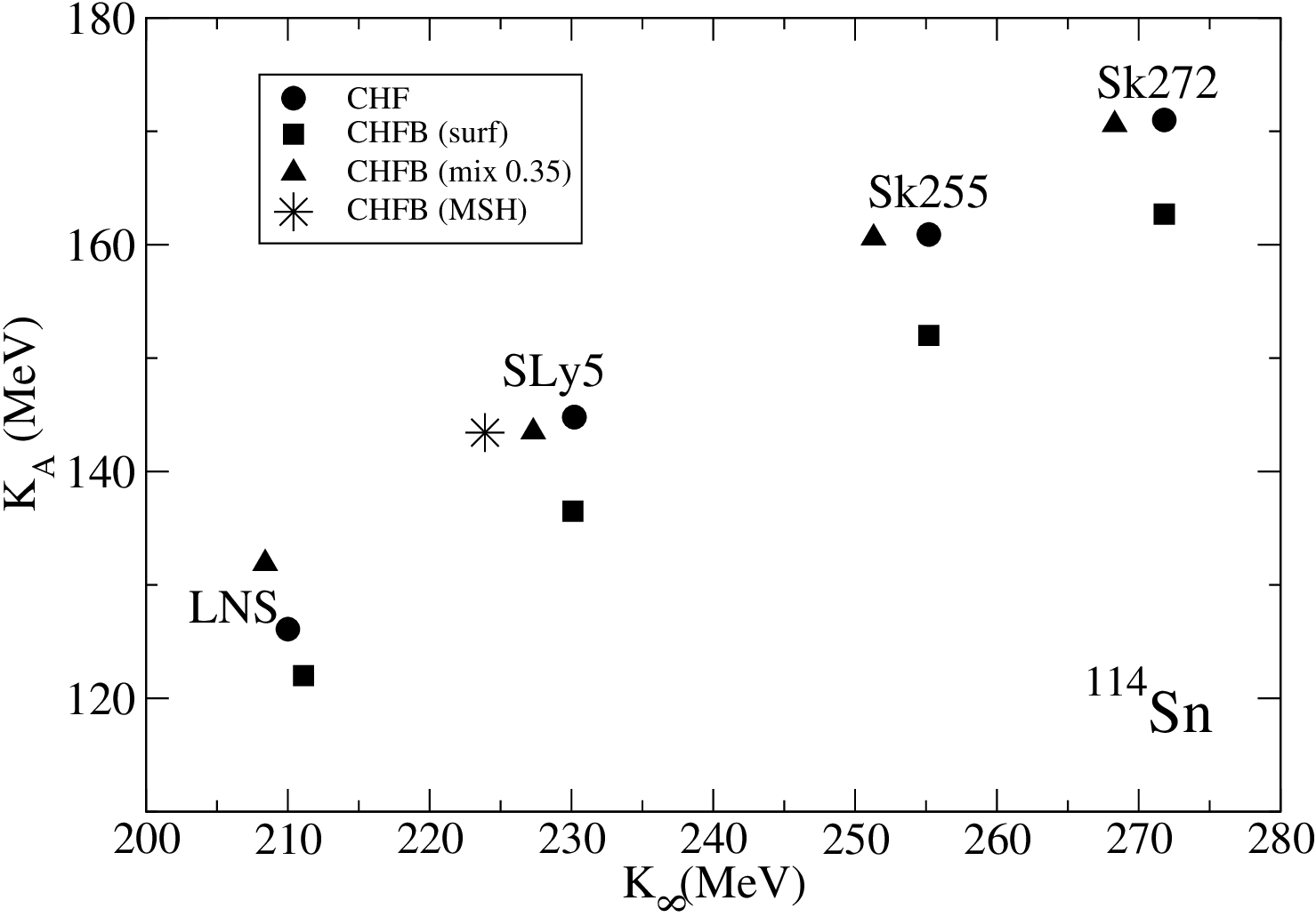}}
\caption{$K_\infty$ versus $K_A$ 
for $^{114}$Sn obtained by the CHF and the 
CHFB method with surface-type and mixed-type pairing interactions for several Skyrme interactions.}
\label{fig:kinfkasn1}
\end{center}
\end{figure}

\begin{figure}[tb]
\begin{center}
\scalebox{0.35}{\includegraphics{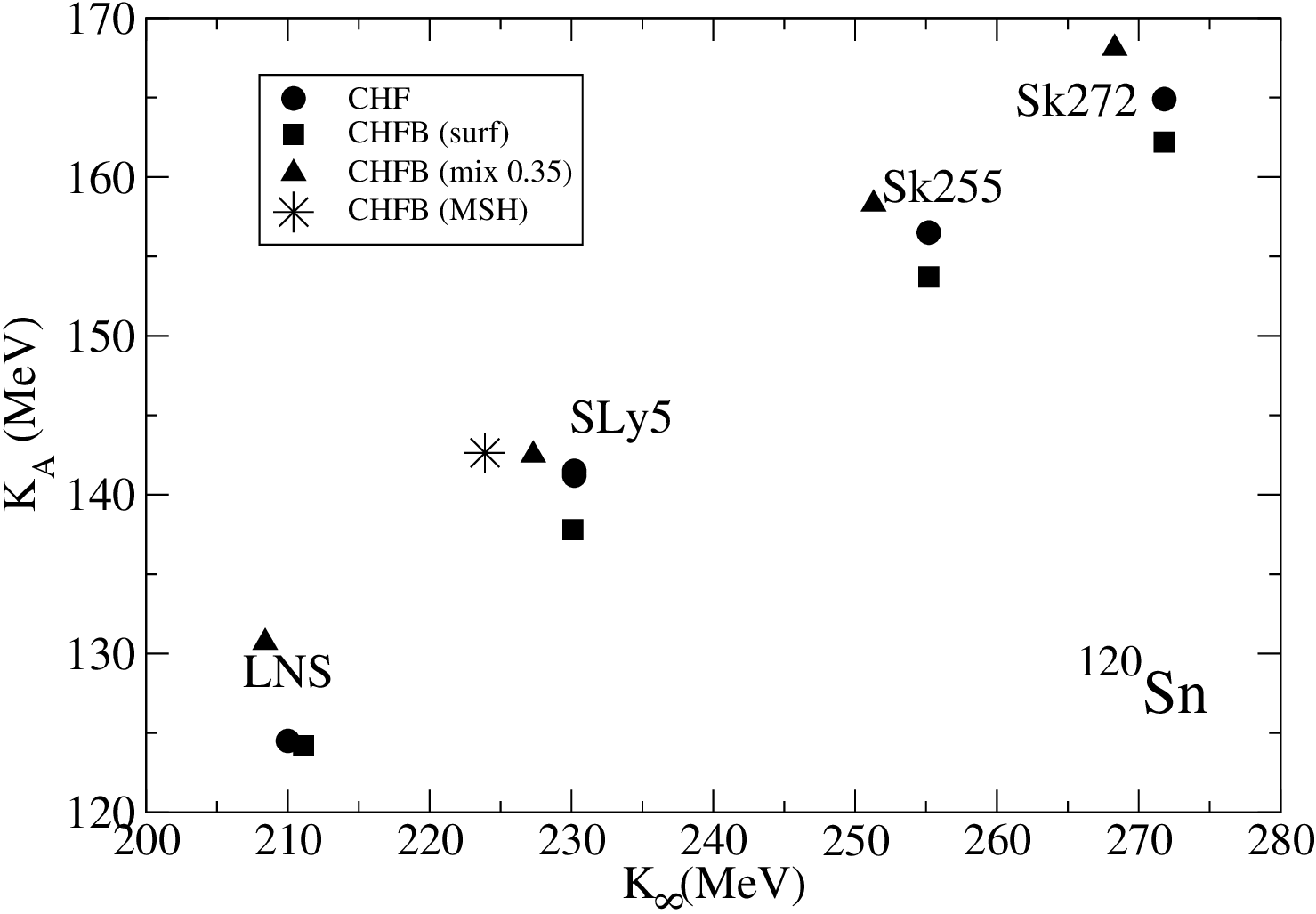}}
\caption{
Same as Fig. 5, but for $^{120}$Sn.}
\label{fig:kinfkasn2}
\end{center}
\end{figure}

It should be stressed that both $K_A$ and $K_{\infty}$ are here evaluated
consistently in a microscopic model. As expected, $K_A$ is clearly
correlated with $K_\infty$.  To find out the pairing effects, we also study
the case of the open-shell nuclei $^{114}$Sn and $^{120}$Sn in Figs.
\ref{fig:kinfkasn1} and \ref{fig:kinfkasn2}, respectively. For each Skyrme
interaction, three results are shown: (i) the CHF result, without pairing,
(ii) the CHFB result using the surface-type pairing interaction, and (iii)
the CHFB result using the mixed-type ($\eta$=0.35) pairing interaction. In
the case of SLy5, the IS+IV MSH pairing interaction is also used. The
surface-type interaction decreases the finite nucleus incompressibility
$K_A$ by about 10\% in $^{114}$Sn and 5\% in $^{120}$Sn whereas the
mixed-type pairing interaction has a negligible effect on $K_A$. Conversely,
it should be reminded that the mixed pairing interaction has some effect on
$K_{\infty}$ whereas the pure surface pairing interaction has a negligible
effect on $K_{\infty}$, as seen on Fig. \ref{fig:kinfkasn1}. In the case of
LNS, the reduction of $K_A$ is smaller for the surface pairing, and an
increase of $K_A$ is even observed for the mixed pairing case. In the SLy5
case, predictions using the IS+IV MSH pairing interactions show no variation
of $K_A$ but affects $K_{\infty}$.

To study how these conclusions in $^{114}$Sn are 
 sensitive to the nuclear shell structure, the results for $^{120}$Sn
  are displayed in Fig. \ref{fig:kinfkasn2}, where
pairing effects are expected to be smaller than in $^{114}$Sn due to the
subshell closure. In this case the reduction of $K_A$ due to the
surface-type pairing effect drops to 5\%. For the mixed pairing interaction
a small increase of $K_A$ is observed. This feature is again more pronounced
in the LNS case. In the SLy5 case, predictions using the IS+IV MSH pairing
interactions show no variation of $K_A$ but affects $K_{\infty}$. It should
be noted that similar trends are observed with the LDA predictions (See
Table \ref{tab:120Sn}). Also, they are consistent with previous studies
\cite{jli08,khan09}. To further study shell effects, the same calculations
have been performed on $^{126}$Sn, and a similar pattern than Fig.
\ref{fig:kinfkasn2} is found, showing that the present results a rather
independent from shell effects in open shell nuclei.

Evidently, the pairing effects tend to decrease the finite nucleus
incompressibility $K_A$ in the surface pairing case, whereas $K_{\infty}$ is
decreased in the mixed pairing case. Hence the question of constraining the
pairing interaction through precise (typical resolution of few hundreds of
keV) GMR measurement is raised. The systematic softness of open-shell Sn and
Cd isotopes measured recently \cite{tli07,tli10,gar09} through the energy of the
GMR might be the sign of a surface pairing interaction at work
\cite{khan09}.

\section{Conclusions}

The effect of superfluidity on the incompressibility has been studied in
both nuclear matter and finite nuclei, using various pairing functionals. A
small effect is observed on the nuclear matter incompressibility at the
saturation density and the symmetry energy, but non-negligible on $L$ and
$K_\mathrm{sym}$.

However at lower density, the pairing effect on the incompressibility is 
significant and can have an substantial impact on neutron stars studies or
on the interpretation of multifragmentation data. It has been shown that the
LDA provides a relevant framework for a qualitative understanding and
interpretation of the microscopic results. The effect of the pairing
correlations is localized near the surface of nuclei and the effect of the
pairing correlations is to make slightly softer nuclear EOS. Especially in
the low density region in nuclear matter, the pairing effect is more
noticeable. This may explain why such effects are expected to happen in the
surface of the finite nuclei. In the case of the IS+IV pairing interaction,
no strong effect is observed on $K_A$. In general, the pairing effects on
the finite nucleus incompressibility $K_A$ are more important when the
interaction is more surface type (larger $\eta$ value). 

This study shows that with respect to current experimental uncertainties,
the pairing effects should be considered when extracting the incompressibility
value from GMR data which can now reach an accuracy of several hundreds of
keV \cite{tli07}. Experimentally it would be useful to measure the GMR on
isotopic chains, including both open-shell and doubly magic nuclei such as
$^{132}$Sn. Such measurements are starting to be undertaken
\cite{tli07,tli10,gar09} and will be extended to unstable nuclei \cite{mon08}.

\section*{Acknowledgement}

This work was supported by the Japanese
Ministry of Education, Culture, Sports, Science and Technology
by Grant-in-Aid for Scientific Research under
the program numbers 20540277 and 22540262, by the ANR NExEN contract,
and by CompStar, a Research Networking Programme of the European Science Foundation.

\end{document}